\documentclass[conference]{./style/IEEEtran2/IEEEtran}
\usepackage{cite}
\pdfoutput=1
\ifCLASSINFOpdf
\else
\fi
\usepackage{amsmath}
\usepackage{listings}
\usepackage{adjustbox}
\usepackage{xspace} % to allow the definition of new commands and nice spaces after them
\usepackage{graphicx}
\usepackage{fancyvrb}
\usepackage{float}
\usepackage{pdfpages}
\usepackage[english]{babel}
\usepackage{multirow}% http://ctan.org/pkg/multirow
\usepackage{array}
\usepackage{color}
\usepackage{soul}
\usepackage{url}
\usepackage{hyperref}
\usepackage{colortbl}
\usepackage[font=footnotesize]{caption}
\usepackage{subcaption}
\usepackage{setspace}
\usepackage{./style/slashbox/slashbox}
\usepackage{./style/bibspacing} % for reducing the space between references
\newcommand{\flo}[1]{{\color{black}#1}}
\newcommand{\sia}[1]{{\color{black}#1}}

\begin{document}

\bstctlcite{IEEEexample:BSTcontrol}

\title{Assessing Data Usefulness for Failure Analysis \\in Anonymized System Logs}
%%\author{[omitted for blind review]}

\author{
\IEEEauthorblockN{Siavash Ghiasvand\IEEEauthorrefmark{2} and Florina M. Ciorba\IEEEauthorrefmark{1}}
\IEEEauthorblockA{
\IEEEauthorrefmark{2}Technische Universit\"at Dresden, Germany\\
\IEEEauthorrefmark{1}University of Basel, Switzerland
}
}

\makeatletter
\def\paragraph{\@startsection{paragraph}%name of the section command
				 	{5}%level/depth of the section command; changing it lower, adds numbers.
					{\z@}% heading indent \z@ - zero points
	                                 {3.25ex \@plus1ex \@minus.2ex}% before skip
        		                        {\z@}% after skip
                  		       {\normalfont\normalsize\bfseries}%style
		       }
\makeatother
\maketitle
\begin{abstract}

System logs are a valuable source of information for the analysis and understanding of systems behavior for the purpose of improving their performance.
Such logs contain various types of information, including sensitive information.
Information deemed sensitive can either directly be extracted from system log entries by correlation of several log entries,
or can be inferred from the combination of the (non-sensitive) information contained within system logs with other logs and/or additional datasets.
The analysis of system logs containing sensitive information compromises data privacy.
Therefore, various anonymization techniques, such as generalization and suppression have been employed, over the years, by data and computing centers to protect the privacy of their users, their data, and the system as a whole.
Privacy-preserving data resulting from anonymization via generalization and suppression may lead to significantly decreased data usefulness, thus, hindering the intended analysis for understanding the system behavior.
Maintaining a balance between data usefulness and privacy preservation, therefore, remains an open and important challenge.
Irreversible encoding of system logs using collision-resistant hashing algorithms, such as \mbox{SHAKE-128}, is a novel approach previously introduced by the authors to mitigate data privacy concerns.
The present work describes a study of the applicability of the encoding approach from earlier work on the system logs of a production high performance computing system.
Moreover, a metric is introduced to assess the data usefulness of the anonymized system logs to detect and identify the failures encountered in the system.

\end{abstract}

\IEEEpeerreviewmaketitle

% !TEX root =  ispdc2018.tex
\section{Introduction}
\label{sec:introduction}

{\em System logs} (or syslogs\footnote{\emph{System log} and \emph{syslog} are used interchangeably in this work}) provide a wide range of information about the behavior of the underlying hardware, software, and users.
\sia{In computing systems, the system-level logging is done by the underlying operating system.
Since November 2015, all TOP500\footnote{https://top500.org/list/2017/11/} high performance computing (HPC) systems are powered by Unix-like operating systems.
Among them, Linux holds the highest share.
The Linux logging protocol is implemented according to RFC 5424~\cite{syslog}, which provides great consistency among system log entries generated by TOP500 HPC systems.}
Part of the syslog information is sensitive and requires protection in order to preserve the privacy of a data subject (i.e., users and systems). 
\flo{Information is deemed sensitive if it can be used to identify a data subject, for example usernames.}
Public disclosure of syslogs (via publishing and sharing) as well as the outsourcing of the syslog analysis to third party stakeholders, is only possible upon providing the required level of data protection and privacy.
\sia{The general data protection regulation (GDPR) of the European Union mandates a Europe-wide personal data privacy protection and will go into force on May 25, 2018~\cite{GDPR}.}
Anonymization is one of the de-facto approaches to provide privacy and protection of data subjects.
The process of log anonymization removes or alters the sensitive data in system logs.
Post-anonymization, the value (or usefulness) of the data may significantly be reduced. 
Yet preserving a high data value within system logs by decreasing the anonymization degree may put data privacy at risk.
The trade-off between data \emph{privacy}  and data \emph{usefulness} plays a critical role in the selection of the anonymization technique~\cite{GDPR2}.

A definition of data usefulness has been proposed by Loukides and Shao \cite{Loukides2007}: data usefulness is \emph{``[t]he extent to which an anonymised [data] table allows required analyses or queries to be made."}
Assessing data usefulness is a non‑trivial challenge, as the nature of data usefulness depends on the intended \emph{usage} of the data.
This work is concerned with assessing the {\em usefulness of data} in anonymized system logs for the purpose of analyzing failures in high performance computing (HPC) systems.

Nowadays, failures in high performance computers (HPC) became the norm rather than the exception.
In the near future, the mean time between failures (MTBF) of HPC systems is expected to be too short, and that current failure recovery mechanisms e.g., checkpoint-restart, will no longer be able to recover the systems from failures.
Early failure detection is a new class of failure recovery methods that can be beneficial for HPC systems with short MTBF.
Detecting failures in their early stage can reduce their negative effects by preventing their propagation to other parts of the system~\cite{Ghaisvand2016a}.

In earlier work,  the authors introduced a system log anonymization approach based on irreversible encoding of log entries via a collision resistant hashing algorithm~\cite{Ghiasvand2018}.
The SHAKE-128~\cite{Shake128} encoding-based anonymization approach guarantees privacy of sensitive data related to users and reduces the required log storage capacity.
\autoref{tab:raw-to-encode} and~\autoref{fig:flowchart} (in Section~\ref{sec:encoding}) provide an overview of the proposed anonymization approach~\cite{Ghiasvand2018}.
Extending prior work~\cite{Ghiasvand2018}, in the present work a study of the applicability of this encoding approach on system logs of Taurus~\cite{Taurus}, a production HPC system at the Technical University of Dresden, Germany is conducted.
Moreover, a metric is introduced to assess the usefulness of data within system logs during and post-encoding for the purpose of detecting and identifying failures encountered in the HPC system.
\flo{The use of the proposed metric on the considered system logs show that the anonymized data retains a high degree of usefulness for the intended purpose of failure detection and identification.}

\sia{The novelty and contributions of the anonymization approach to retain data usefulness proposed in this work consists of: 
(1)~anonymization of \emph{unstructured} system log messages;  
(2)~preservation of the usefulness of system logs (especially for semi-/automated failure analysis); 
(3)~a \emph{guarantee} for protecting data privacy;
(4)~significant reduction of the required capacity for the storage of the anonymized system logs;
(5)~offering the choice to set the degree of generalization according to the intended data usage; and 
(6)~providing readily analyzable data, anonymized and encoded, that does not require decoding before analysis.
}

The remainder of this work is structured as follows.
Section~\ref{sec:relatedwork} overviews the work related to data anonymization and usefulness.
The anonymization approach employed herein is described in Section~\ref{sec:encoding}. 
The proposed data usefulness metric is given in Section~\ref{the-usefulness}, while a quantitative evaluation of the data usefulness is provided in Section~\ref{sec:evaluation}.
The paper concludes and introduces important future work directions in Section~\ref{sec:conclusion}.

% !TEX root =  ispdc2018.tex

\section{Related work}
\label{sec:relatedwork}

Generalization and suppression are two well-known methods for data anonymization.
These methods either group or remove data, in order to reduce uniqueness, and thus, the chance of identification of individual data subjects from the records in the dataset.
The $k$-anonymity protection model~\cite{Sweeney2002} has been introduced as a model for protecting privacy.
\sia{Although $k$-anonymity addresses the main challenge of data privacy in anonymized datasets, it has several shortcomings (e.g.,~attribute disclosure, complementary data release)}. 
To overcome these shortcomings, several models such as $l$-diversity~\cite{Machanavajjhala2006} and $t$-closeness~\cite{Li2007} have been introduced.
\sia{These models reduce the data representation granularity (grouping) beyond the level used in $k$-anonymity, which can result in decreased data usefulness.}
Recent studies~\cite{Kumar2018} considered an integration of both, the $l$-diversity and $t$-closeness models.
To the authors' best knowledge, no anonymization model can guarantee full privacy of data contained in anonymized form in the given datasets.

\sia{According to the articles 2, 4(1) and(5) and recitals (14), (15), (26), (27), (29) and (30) of the GDPR~\cite{GDPR}, in order to analyze sensitive information, an irreversible anonymization of personal data must be guaranteed.}

The process of data anonymization incurs a certain degree of information loss.
With significant information loss comes decreased usefulness of the anonymized data.
Various studies attempted to address the problem of achieving $k$-anonymity protection with minimal information loss.
Gionis and Tassa proved that solving the problem for the two conflicting goals above is NP-hard~\cite{Gionis2009}.
Later, it has been shown that dynamic optimization of the anonymization process considerably reduces the loss of information~\cite{Murakami2017}. 
In another attempt to address the high information loss during data anonymization, \emph{utility-based anonymization} methods were proposed.
Xu {\em et al.}~\cite{Xu2006} introduced an approach which first, specifies the utility of each attribute, and second, proposes two heuristic \emph{local recording}-based anonymization methods~\cite{Terrovitis2011} to boost the quality of the analysis later.
A \emph{data relocation} mechanism has also been applied to reduce granularity and populate small groups of tuples to increase data usefulness~\cite{Nergiz2013}.
In another similar effort, quasi-identifiers have been divided into two groups of ordered and unordered attributes~\cite{wang2010}. 
To reduce the information loss, more flexible strategies for data generalization have been applied on the unordered attributes.
More recently, \emph{co-utility}~\cite{Soria2018} has been introduced as a global distributed mechanism for data anonymization, such that a balance between data utility and data privacy is achieved.
Although these efforts decrease the information loss during data anonymization, they still cannot guarantee data privacy.
The non-zero probability of privacy breaches through anonymization by the use of approaches such as those mentioned above has experimentally been determined~\cite{Wong2011}.

\sia{Quantifying data utility\footnote{\emph{Utility} and \emph{usefulness} are used interchangeably in this work.}, which is a qualitative property of data, provides a measure to control the balance between privacy and utility of data.}
A number of studies proposed such measures to quantify the utility of anonymized data.
%\flo{The previous sentence is unclear. It is like saying: umbrellas have been invented, but we have never heard of rain; therefore, umbrellas seem irrelevant. Need to write a sentence before the previous one, to state the need for quantification of the data usefulness (or utility - still need to add that both are used interchangeably and that both mean the same thing!).}
\sia{Data utility is mostly described as the amount of information loss.}
Information loss, in general, can be quantified according to the uncertain change in attribute values during the anonymization~\cite{wang2010}, via \mbox{result-driven} approaches to compare the data before and after anonymization~\cite{Templ2017}, or even according to the data entropy in the dataset~\cite{Loukides2007}.
These measures are generally divided into two categories: 
(1)~\emph{entropy-based} and 
(2)~\emph{distance-based} (e.g., the Hellinger distance).
Furthermore, most of the above mentioned usefulness quantification approaches are implemented into data anonymization tools, such as ARX~\cite{Prasser2015}.

All existing approaches for quantification of data usefulness aim to increase data privacy, data utility, or both, in anonymized datasets.
However, all approaches implicitly make the fundamental assumption of having a structured format for the data entries. 
In reality, system log entries are of mixed \emph{structured} and \emph{unstructured} data formats.
The structured part contains the meta-data (e.g., time or location related to the particular syslog entry) and the unstructured part contains the detailed event information.
Sensitive information mainly resides within the unstructured part of the data.
In the authors' best knowledge, due to the unstructured nature of the detailed event information (no two distinct events generate the same information pattern), none of the existing approaches provide a {\em utility-based anonymization} of system logs. 
\sia{Moreover, although a few studies addressed the de-identification of unstructured datasets~\cite{GARDNER2009}, none of these studies, nor the known privacy models (e.g., $k$-anonymity) guarantee data privacy.} 

Encryption is a viable alternative to the above methods to provide certain degrees of data privacy without any information loss.
\sia{However, encryption is reversible, the encryption key having to be \flo{securely preserved yet} also shared in order to make further analysis possible.}
Therefore, encryption can only be used within a trusted environment.
\sia{Any form of encryption is theoretically breakable, \flo{provided enough time and computational power.}}
Therefore, log encryption is \flo{also} not a suitable approach for sharing or publishing system logs.

% !TEX root =  ispdc2018.tex
\section{Privacy-Preserving Anonymization\\ of System Logs}
\label{sec:encoding}

System log entries typically consist of three parts, (1)~\emph{timestamp}, (2)~\emph{source}, and (3)~\emph{message}.
\sia{Timestamp and source contain structured data while message is unstructured.}
The timestamp denotes the time at which an event occurs.
The source provides information about the location of the event occurrence, while the message describes the event properties.
\autoref{tab:raw-to-encode-1} illustrates three syslog entries divided into these three main parts. %
In this example, the timestamps are in the UNIX time format, the sources are node IDs, and the messages contain event details.

\begin{figure}
	\begin{center}
		\includegraphics[width=.45\textwidth]{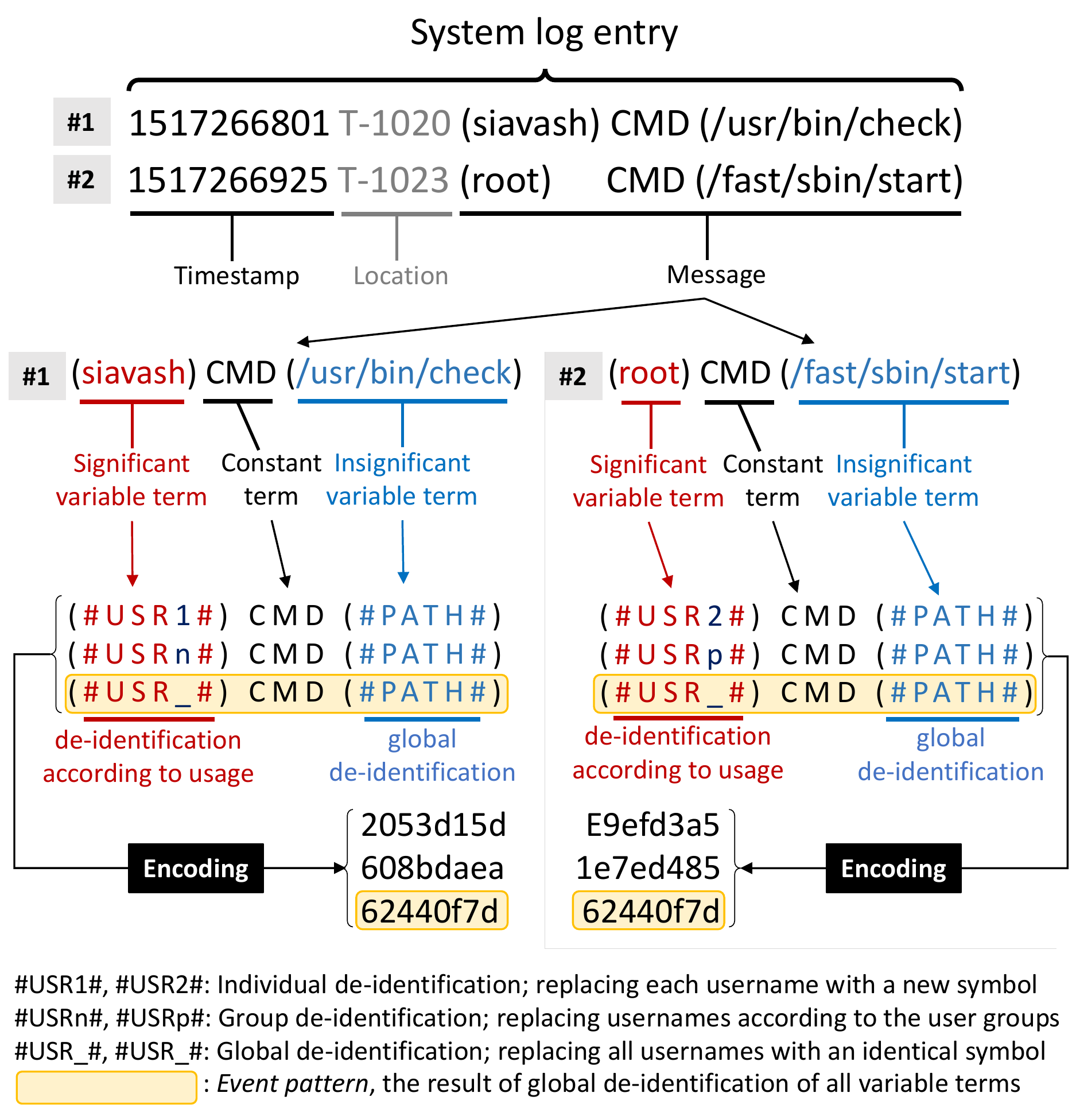}
		\caption{The \emph{event pattern} is the result of a full de-identification of syslog entires. 
		Note that other forms of de-identification are also possible. 
		The final encoding step guarantees full data privacy.}
		\label{fig:flowchart}
	\end{center}
	\vspace{-0.2cm}
\end{figure}

Earlier work introduced $PaRS$\footnote{Anonymization of Syslogs for Preserving \textbf{\underline{P}}rivacy \textbf{\underline{a}}nd \textbf{\underline{R}}educing \textbf{\underline{S}}torage.} as an anonymization approach which employs an irreversible encoding method to guarantee the full anonymization of system logs~\cite{Ghiasvand2018}.
\flo{The $PaRS$ workflow is illustrated in~\autoref{fig:flowchart} which also summarizes the necessary terminology.}
$PaRS$ consists of two main steps:~de-identification and~encoding.
The \emph{de-identification} step replaces the message part of each syslog entry with its relevant \emph{event pattern} as shown in~\autoref{tab:raw-to-encode-2}.
The \emph{encoding} step encodes the event patterns, from the de-identification step, via an irreversible collision-resistant hashing algorithm, namely SHAKE-128, into hash keys, as shown in~\autoref{tab:raw-to-encode-3}.
The \emph{timestamp} and \emph{source} remain \flo{unchanged}.

\begin{table}[!t]
	\fontsize{6pt}{10pt}\selectfont
	\caption{Anonymization of syslog entries via $PaRS$}
	\centering
	\begin{subtable}[t]{.45\textwidth}
		\caption{Raw system log entries}
		\label{tab:raw-to-encode-1}
		\begin{tabular}{l|l|l}
		%\hline\noalign{\smallskip}
		\multicolumn{1}{p{1cm}}{\textbf{Timestamp}} &
		\multicolumn{1}{p{0.6cm}}{\textbf{Source}} &
		\multicolumn{1}{p{5.6cm}}{\textbf{Message}}\\
		\hline\noalign{\smallskip}
		\verb|1515625261|&\verb|T-1230|&\verb|(siavash) CMD (/home/config.sh > output.stat)|\\
		\verb|1515625370|&\verb|T-3417|&\verb|pam_unix: session closed for siavash|\\
		\verb|1515625713|&\verb|T-6201|&\verb|disabling lock debugging due to kernel taint|\\
		\end{tabular}
	\end{subtable}
	\begin{subtable}[t]{.45\textwidth}
		\caption{Event patterns}
		\label{tab:raw-to-encode-2}
		\begin{tabular}{l|l|l}
			\multicolumn{1}{p{1cm}}{\textbf{Timestamp}} &
			\multicolumn{1}{p{0.6cm}}{\textbf{Source}} &
			\multicolumn{1}{p{5.6cm}}{\textbf{Message}}\\
			\hline\noalign{\smallskip}
			\verb|1515625261|&\verb|T-1230|&\verb|(#USER#) CMD (#PATH# > #PATH#)|\\
			\verb|1515625370|&\verb|T-3417|&\verb|pam_unix: session closed for #USER#|\\
			\verb|1515625713|&\verb|T-6201|&\verb|disabling lock debugging due to kernel taint|\\		
		\end{tabular}
	\end{subtable}

	\begin{subtable}[t]{.45\textwidth}
		\caption{Encoded system log entries}
		\label{tab:raw-to-encode-3}
		\begin{tabular}{l|l|l}
			\multicolumn{1}{p{1cm}}{\textbf{Timestamp}} &
			\multicolumn{1}{p{0.6cm}}{\textbf{Source}} &
			\multicolumn{1}{p{5.6cm}}{\textbf{Message}}\\
			\hline\noalign{\smallskip}
			\verb|1515625261|&\verb|T-1230|&\verb|1808e388|\\ %1808e388919ee30bfaf7fdc5aed8823d
			\verb|1515625370|&\verb|T-3417|&\verb|0964de42|\\%0964de42ac897ed285eea6e054fb6e90
			\verb|1515625713|&\verb|T-6201|&\verb|59f2da35|\\%59f2da35bcc39525b87932b4cc1f3d68
		\end{tabular}
	\end{subtable}
	\label{tab:raw-to-encode}
\end{table}

\flo{The message parts of syslog entries consist of one or more terms.}
 A \emph{term} is a string of characters with a certain semantic (e.g., root, 2, CMD).
Each term can either be \emph{constant} or \emph{variable}.
A constant term remains identical in all syslog entries.
In contrast, a variable term can take different values across different syslog entries.
\flo{Variable terms can be classified into \emph{significant} and \emph{insignificant} (variable) terms, depending of the intended future \emph{usage} of the anonymized logs.}
The significant variable terms can be de-identified in various forms according to the intended usage.
In contrast, all insignificant variable terms, \flo{for example, /usr/bin/} must be de-identified, via substitution by an identical symbol, such as \#PATH\#.
\flo{This form of de-identification, which applies to insignificant terms by default, is referred to as \emph{global} de-identification, while the substitution of each term by an individual symbol is called \emph{individual} de-identification.}
\flo{Global de-identification provides the highest degree of generalization, while individual de-identification prevents any generalization.}

Categorizing terms into groups and substituting all terms of each group with an identical group symbol is called \emph{group} de-identification.
The group de-identification can provide various degrees of generalization according to the grouping granularity.
Consider the sample log entry in \autoref{lst:singleLogEntry}.
This log entry describes that user \verb|siavash| executed the command \verb|/usr/bin/check| on computer \verb|T-1020| \flo{on} \verb|January 29, 2018 11:00:01 PM|.
In this entry \verb|siavash| and \verb|/usr/bin/check| are variable terms, and \verb|CMD| is a constant term.
\flo{With} respect to the intended future usage of the anonymized system logs in this work, namely failure analysis, \verb|siavash| is considered as a significant variable term and \verb|/usr/bin/check| as an insignificant variable term.

\begin{lstlisting}[title=SS,language=C,caption=Sample log  entry,label=lst:singleLogEntry,frame=none,basicstyle=\scriptsize\ttfamily,keepspaces=true]
1517266801 T-1020 (siavash) CMD (/usr/bin/check)
\end{lstlisting}

The event pattern of a syslog entry is generated through global de-identification of all variable terms in the message part of the respective syslog entry.
Therefore, transforming log entries into event patterns at first, and, subsequently, encoding the event patterns into hash keys, suppresses all potential sensitive information within the log entries. 
Moreover, the \emph{similarity} between syslog entires is preserved such that the results of further data analysis are not skewed.
\flo{Although a hash key might appear devoid of semantic, given the \mbox{one-to-one} relation between hash keys and event patterns, it is always possible to reaccredit the original semantic to the pattern denoted by a hash key.}
This accreditation can only be done by the owners of the adequate information about the event patterns and the hashing function.
\sia{However, regardless of the reaccreditation of the original semantic to the pattern, it is always possible to track similar events according to the similarity of their event patterns.}

The final output of $PaRS$ consists of the timestamp and source of the log entries in their original raw format and the anonymized message part as a hash key.
Depending on the strictness of the applicable privacy guidelines, the \emph{meaning} of each hash key may also be added to the final output.
\autoref{tab:hash-meaning} contains the final output of applying $PaRS$ on the given sample syslog entries in~\autoref{tab:raw-to-encode-1}, accompanied by the hash key meanings.

\begin{table}[!t]
	\fontsize{6pt}{10pt}\selectfont
	\caption{Final output of $PaRS$}
	\centering
	\begin{tabular}{l|l|l|l}
		\textbf{Timestamp}& \textbf{Source}& \textbf{Hash key}& \textbf{The meaning}\\
		\hline
		\verb|1515625261|&\verb|T-1230|&\verb|1808e388|&\Verb|A command executed by a user|\\
		\verb|1515625370|&\verb|T-3417|&\verb|0964de42|&\Verb|A user logged out|\\
		\verb|1515625713|&\verb|T-6201|&\verb|59f2da35|&\Verb|Kernel is in taint mode|\\
	\end{tabular}
	%	\subcaption{Hash key reference table}
	\label{tab:hash-meaning}
\end{table}

% !TEX root =  ispdc2018.tex
\subsection{Usage Oriented Processing of System Logs}
\label{sec:usefulness}

Conducting any form of failure \flo{detection and identification, for the purpose of failure analysis, requires in-depth details} about the actual state of the system; system logs readily contain such information.
The usage of an HPC system is regulated by the privacy guidelines in force, according to its functionality, production environment, and administration domain.
Depending on the applicable privacy regulations, certain information within the system logs may be considered as sensitive information.
Examples include usernames and IP addresses.
Information deemed sensitive on one HPC system may be deemed not sensitive on a different HPC system.
Analyzing and publishing raw system logs, that may contain sensitive information endangers the privacy of data subjects such as users, system owners, system vendors, and others.
Therefore, data anonymization is required \emph{before} the analysis and sharing of (raw) system logs with such sensitive information.

The data protection and privacy guidelines of each system mandate the removal of a certain sensitive information from system logs.
Therefore, a certain amount of information loss cannot be prevented during the anonymization phase.
Post-anonymization, the system logs may have already lost their usefulness for certain types of analyses.
For example, the anonymized system logs of a computing system, with a privacy guideline that mandates complete removal of all usernames from system logs before any analysis, are not useful for analyses performed for user accounting purposes.

Terms within system log entries can be divided into two groups:
(1)~\emph{quasi-identifiers}, and (2)~\emph{sensitive-attributes}, depending on their sensitiveness.
Quasi-identifiers are the key values used to identify certain syslog entries and can be shared within the public domain (e.g., timestamp).
The sensitive-attributes must always remain confidential (e.g., username).

In a study by Loukides and Shao~\cite{Loukides2007} data usefulness and data privacy are evaluated via $k$-anonymity according to the diversity of the quasi-identifiers and the sensitive-attributes.
The usefulness of data is defined as the average of \emph{tuple diversity} among different groups of tuples.
The tuple diversity, in turn, is defined as the summation of \emph{attribute diversity} between the quasi-identifiers of each tuple.

\flo{The usefulness metric proposed in this work (refer to Equation~(1) in Section~\ref{the-usefulness}) is inspired by the study of Loukides and Shao~\cite{Loukides2007}.
Other aspects of their study are not relevant to this work.
The anonymization approach employed herein, $PaRS$~\cite{Ghiasvand2018}, guarantees data privacy.
Employing the hashing algorithm in $PaRS$, encapsulates all terms into a single hash key which prevents any accidental leakage of sensitive information into anonymized system logs.
Therefore, the system logs anonymized via $PaRS$ do not encounter any of the privacy challenges encountered by those anonymized based on \emph{$k$-anonymity} or \emph{differential privacy}~\cite{Dwork2006} data protection models.}
As illustrated in~\autoref{tab:hash-meaning}, the three parts of each syslog entry (timestamp, source, and message), after anonymization, become quasi-identifiers and they may be freely shared in the public domain.
Finally, applying $PaRS$ on system log entries also provides a \flo{significant gains in} the required storage space.

\autoref{tab:notation} summarizes the notation used in this work. this notation is further detailed in Section~\ref{the-usefulness}.

\begin{table}[!t]
	\linespread{0.9}
	\fontsize{6.9pt}{10pt}\selectfont
	\caption{Notation employed in this work}
	\centering
	\begin{tabular}{l|l}
		\textbf{Notation} & \textbf{Definition} \\
		\hline
		$e_r$			& Raw log entry \\
		$e_d$			& De-identified log entry \\
		$v_i$			& Significant variable term within a raw log entry $e_r$, $0 \leq i \leq n$ \\
		$N^{val}_{v_i}$		& Number of values for a significant variable term $v_i$ in a raw log entry $e_r$ \\

		$N^{s}_{v_i}$		& Number of de-identification symbols that replace the  \\
		& Significant variable term $v_i$ in a \mbox{de-identified} log entry $e_d$ \\
		& \hspace{0.5cm} Individual de-identification $N^{s}_{v_i}=N^{val}_{v_i}$  \\
		& \hspace{0.5cm} Group de-identification $N^{s}_{v_i}<N^{val}_{v_i}$   \\
		& \hspace{0.5cm} Global de-identification $N^{s}_{v_i}=1$     \\
		$N_{e}$			& Number of raw log entries $e_r$ in a day \\
		$f_{p}$			& Frequency of event pattern $p$ in a collection of log entries in a day \\
		$D_p$			& Degree of dominance of an event pattern $p_i$ in a collection of   \\
		& encoded system log entries in a day \\
		& \hspace{0.5cm}  $D_{p}=f_p/N_e$ \\ 
		$U$				& Data usefulness
		
	\end{tabular}
	\label{tab:notation}
\end{table}

\subsection{Categorization of System Logs}
\label{subsec:event-patterns}

The \emph{event pattern} of a system log is obtained via global de-identification of all variable terms in message part of the syslog entry.
\flo{By} differentiating between similar variable terms, \mbox{less-strict} forms of de-identification can be applied.
%\footnote{\flo{Via is never used to start a sentence, but typically only within sentences.}}
\autoref{tab:event-patterns} illustrates the message part of a sample raw syslog entry and four different degrees de-identification that can be applied to it.
In this example, the terms \verb|siavash|, \verb|/home/siavash/config.sh|, and \verb|/dev/null| are variable terms.
All other symbols are constant terms.
In de-identifications of \emph{type 1} and \emph{type 2}, various paths are differentiated, while the username is replaced with its dedicated constant identifier.

\begin{table}[!t]
	\linespread{0.9}
	\fontsize{6pt}{10pt}\selectfont
	\caption{De-identification of raw system log entries}
	\centering
	\begin{tabular}{l l@{} l@{}}
		De-id. degree & Message & Hash key\\
		\hline\noalign{\smallskip}
		Raw&\verb|(siavash) cmd (/home/siavash/config.sh > /dev/null)|&12e1577b\\
		Global &\verb|(#USR_#) cmd (#PATH# > #PATH#)|&60b57133\\
		Type 1&\verb|(#USR_#) cmd (#PATH1# > #PATH2#)|&0479abde\\
		Type 2&\verb|(#USR1#) cmd (#PATH1# > #PATH2#)|&e78d2b56\\
		Type 3&\verb|(#USR1#) cmd (#PATH# > #PATH#)|&d4ad931b\\

	\end{tabular}
	\label{tab:event-patterns}
	%	\vspace{5 mm}
\end{table}

\flo{With more differentiations between variable terms, a larger number of hash keys will need to be generated.}
Each hash key represents a particular \emph{event} with certain properties.
There might be other hash keys which refer to the same event yet with different properties.
All hash keys that represent the same event form an \emph{event category}.
All events in an event category, have identical event patterns, generated via the full global de-identification of their respective syslog entry.

\autoref{tab:event-class-1} shows two sample syslog entries \sia{with different properties} that belong to the same event category.
In~\autoref{tab:event-class-2} the encoded syslog entries (hash keys) with different forms of de-identification are shown.
\begin{table}[!t]
	\linespread{0.9}
	\fontsize{6pt}{10pt}\selectfont
	\caption{Illustration of the effect of de-identification degree}
	\centering
	\begin{subtable}[t]{.45\textwidth}
		\centering
		\caption{Sample raw system log entries}
		\label{tab:event-class-1}
		\begin{tabular}{l l@{} l@{}}
			Entry & Message & Hash key\\
			\hline\noalign{\smallskip}
			Log 1&\verb|(siavash) cmd (/home/siavash/config.sh > /dev/null)|&12e1577b\\
			Log 2&\verb|(florina) cmd (/home/florina/setup.sh > /dev/null)|&0616d4c7\\
		\end{tabular}
	\end{subtable}
	
	\begin{subtable}[t]{.45\textwidth}
		\caption{Encoded log entries, various degrees of de-identification}
		\label{tab:event-class-2}
		\centering
		\begin{tabular}{l|c|c|c}
			De-id. degree & Encoded Log 1 & Encoded Log 2 & Hash key identical?\\
			\hline
			Raw		&12e1577b&0616d4c7&N\\
			Global	&60b57133&60b57133&Y\\
			Type 1	&0479abde&5d4df517&N\\
			Type 2	&e78d2b56&f3a700a7&N\\
			Type 3	&d4ad931b&204679a8&N\\
		\end{tabular}
	\end{subtable}
	\label{tab:event-class}
\end{table}
Not surprisingly, only global de-identification generates identical hash key for both sample log entries in~\autoref{tab:event-class-1}.
To preserve the semantic relation between anonymized system logs after other forms of de-identification, the event pattern of each syslog entry identifies its event category.
\autoref{tab:sample-logs-1} provides a set of sample syslog entries, together with their hash keys and event categories.

\begin{table}[!t]
	\linespread{0.9}
	\fontsize{5.5pt}{10pt}\selectfont
	\caption{Processing of sample system log entries}
	\centering
	\begin{subtable}{.45\textwidth}
		\caption{Raw syslog entries}
		\label{tab:sample-logs-1}
		\centering
		\begin{tabular}{l @{}| l @{} | l | l | l}
			\multicolumn{1}{p{0.1cm}|}{\#}&
			\multicolumn{1}{p{5.0cm}|}{\textbf{Message}} &
			\textbf{Hash key} &
			\textbf{Category} \\
			\hline
			1 & \verb|(siavash) CMD (/usr/bin/check >/dev/null 2>&1)| & \cellcolor{blue!5}a8848910 & \cellcolor{blue!15}66dc2742 \\
			2 & \verb|(florina) CMD (/usr/lib32/lm/lm1 1 1)| & 10a31145 & \cellcolor{blue!15}66dc2742 \\
			3 & \verb|(siavash) CMD (run-parts /etc/cron.hourly)| & \cellcolor{gray!15}a6a420a6 & \cellcolor{blue!15}66dc2742 \\
			4 & \verb|starting 0anacron| & \cellcolor{red!5}47c6b01d & \cellcolor{red!15}dd740712 \\
			5 & \verb|Anacron started on 2018-01-30| & bd94c195 & \cellcolor{green!15}e5a59462 \\
			6 & \verb|Jobs will be executed sequentially| & \cellcolor{yellow!5}f1e7eac3 & \cellcolor{yellow!15}f1e7eac3\\
			7 & \verb|Normal exit (0 jobs run)| & e46c1bdb & \cellcolor{brown!15}eac7924f \\
			8 & \verb|finished 0anacron| & \cellcolor{pink!5}76690e70 & \cellcolor{pink!15}a5803a8a \\
			9 & \verb|(siavash) CMD (/usr/lib32/lm/lm1 1 1)| & bacc6097 & \cellcolor{blue!15}66dc2742 \\
			10& \verb|(root) CMD (/usr/lib32/cl/cl2 1 1)| & eefabc01 & \cellcolor{blue!15}66dc2742 \\
			11& \verb|(root) CMD (/usr/lib64/lm/lm1 1 1)| & 4237ce2c & \cellcolor{blue!15}66dc2742 \\
			12& \verb|(siavash) CMD (/usr/bin/check >/dev/null 2>&1)| & \cellcolor{blue!5}a8848910 & \cellcolor{blue!15}66dc2742 \\
			13& \verb|(florina) CMD (/usr/bin/run >/dev/null 2>&1)| & 8470df87 & \cellcolor{blue!15}66dc2742 \\
			14& \verb|(siavash) CMD (/usr/bin/exec >/dev/null 2>&1)| & dd0e4a50 & \cellcolor{blue!15}66dc2742 \\
			15& \verb|(siavash) CMD (run-parts /etc/cron.hourly)| & \cellcolor{gray!15}a6a420a6 & \cellcolor{blue!15}66dc2742 \\
			16& \verb|starting 0anacron| & \cellcolor{red!5}47c6b01d & \cellcolor{red!15}dd740712 \\
			17& \verb|Anacron started on 2018-01-31| & d414932d & \cellcolor{green!15}e5a59462 \\
			18& \verb|Jobs will be executed sequentially| & \cellcolor{yellow!5}f1e7eac3 & \cellcolor{yellow!15}f1e7eac3 \\
			19& \verb|Normal exit (4 jobs run)| & 0c3b639c & \cellcolor{brown!15}eac7924f \\
			20& \verb|finished 0anacron| & \cellcolor{pink!5}76690e70 & \cellcolor{pink!15}a5803a8a\\	
		\end{tabular}
	\end{subtable}

	\begin{subtable}{.45\textwidth}
		\caption{Pre-anonymized entries (insignificant terms:  de-identified)}
		\label{tab:sample-logs-2}
		\centering
		\begin{tabular}{l @{}| l @{} | l | l | l |}
		%	\noalign{\smallskip}
			\multicolumn{1}{p{0.1cm}|}{\#}&
			\multicolumn{1}{p{3.9cm}|}{\textbf{Message}} &
			\multicolumn{1}{p{0.8cm}|}{\textbf{Significant term}} &
			\textbf{Hash key} &
			\textbf{Category} \\
			\hline
			1 & \verb|(siavash) CMD (#PATH#)| & siavash & \cellcolor{blue!5}bb2d95d2 & \cellcolor{blue!15}66dc2742 \\
			2 & \verb|(florina) CMD (#PATH#)| & florina & \cellcolor{gray!15}23343ad0 & \cellcolor{blue!15}66dc2742 \\
			3 & \verb|(siavash) CMD (#PATH#)| & siavash & \cellcolor{blue!5}bb2d95d2 & \cellcolor{blue!15}66dc2742 \\
			4 & \verb|starting 0anacron| & 0anacron & \cellcolor{red!5}47c6b01d & \cellcolor{red!15}dd740712 \\
			5 & \verb|Anacron started on #TIME#| & Anacron & \cellcolor{green!5}22bb4f1a & \cellcolor{green!15}e5a59462 \\
			6 & \verb|Jobs will be executed sequentially| & - & \cellcolor{yellow!15}f1e7eac3 & \cellcolor{yellow!15}f1e7eac3\\
			7 & \verb|Normal exit (0 jobs run)| & 0 & e46c1bdb & \cellcolor{brown!15}eac7924f \\
			8 & \verb|finished 0anacron| & 0anacron & \cellcolor{pink!5}76690e70 & \cellcolor{pink!15}a5803a8a \\
			9 & \verb|(siavash) CMD (#PATH#)| & siavash & \cellcolor{blue!5}bb2d95d2 & \cellcolor{blue!15}66dc2742 \\
			10& \verb|(root) CMD (#PATH#)| & root & 752d8638 & \cellcolor{blue!15}66dc2742 \\
			11& \verb|(root) CMD (#PATH#)| & root & 752d8638 & \cellcolor{blue!15}66dc2742 \\
			12& \verb|(siavash) CMD (#PATH#)| & siavash & \cellcolor{blue!5}bb2d95d2 & \cellcolor{blue!15}66dc2742 \\
			13& \verb|(florina) CMD (#PATH#)| & florina & \cellcolor{gray!15}23343ad0 & \cellcolor{blue!15}66dc2742 \\
			14& \verb|(siavash) CMD (#PATH#)| & siavash & \cellcolor{blue!5}bb2d95d2 & \cellcolor{blue!15}66dc2742 \\
			15& \verb|(siavash) CMD (#PATH#)| & siavash & \cellcolor{blue!5}bb2d95d2 & \cellcolor{blue!15}66dc2742 \\
			16& \verb|starting 0anacron| & 0anacron & \cellcolor{red!5}47c6b01d & \cellcolor{red!15}dd740712 \\
			17& \verb|Anacron started on #TIME#| & Anacron & \cellcolor{green!5}22bb4f1a & \cellcolor{green!15}e5a59462 \\
			18& \verb|Jobs will be executed sequentially| & - & \cellcolor{yellow!15}f1e7eac3 & \cellcolor{yellow!15}f1e7eac3 \\
			19& \verb|Normal exit (4 jobs run)| & 4 & 0c3b639c & \cellcolor{brown!15}eac7924f \\
			20& \verb|finished 0anacron| & 0anacron & \cellcolor{pink!5}76690e70 & \cellcolor{pink!15}a5803a8a\\	
		\end{tabular}
	\end{subtable}

	\begin{subtable}{.45\textwidth}
		\caption{\emph{Significant} terms per event pattern for different de-identification levels}
		\label{tab:sample-logs-3}
		\centering
		\begin{tabular}{c|l|l|l|c}
			Log 	& Without de-identification /  				& Group 							& Global 							& Degree of \\  
			entry	& Individual de-identification 					& de-identification 				& de-identification				 	& dominance \\
			(range)	& $\prod_{i=1}^{n} N^{val}_{v_i}$ 	& $\prod_{i=1}^{n} N^{s}_{v_i}$	& $\prod_{i=1}^{n} N^{s}_{v_i}$ 	& $D_{p}$ \\
			\hline
			\multicolumn{1}{p{0.8cm}|}{1-3, 9-15} & \textbf{3} (usernames) & 2 (user groups) & 1 & 0.5 \\
			\hline
			4, 16 & \textbf{1} (daemon name)& 1 & 1 & 0.1 \\
			\hline
			5, 17 & \textbf{1} (daemon name)& 1 & 1 & 0.1 \\
			\hline
			6, 18 & No \emph{significant} variables& 1 & 1 & 0.1 \\
			\hline
			7, 19 & \textbf{2} (number of jobs)& 1 & 1 & 0.1 \\
			\hline
			8, 20 & \textbf{1} (daemon name)& 1 & 1 & 0.1 \\
		\end{tabular}
	\end{subtable}
	\label{tab:sample-logs}
\end{table}

\subsection{The Sample Usage: Failure Analysis}
\label{subsec:failure-analysis}
System logs are valuable source of information to study the system behavior in order to detect anomalies and failures.
Series of correlated failures can propagate through a system and form chains of failures.
Early detection of failures in large scale high performance computing systems decreases the potential damages caused by \flo{failure chains}.
The main focus in this work is to provide anonymized syslogs which are useful for the purpose of failure \flo{detection and identification} in HPC systems.
The current failure analysis method used in this work is solely for exemplifying the proposed approach.
The proposed approach is, however, applicable to other types of behavior, other systems, other types of logs, and analysis.

The use case of this study is fault detection and behavior analysis on Taurus~\cite{Taurus}, a production high performance computing system with 2,046 nodes and more than 41,000 cores.
The goal is to \flo{detect and identify} failures on Taurus as \flo{early} as possible.
The source of information for this analysis is a time series of events extracted from Taurus system logs.
The detection mechanism monitors various parameters which include frequency, re-occurrence pattern, absence, sequences, and the time interval between events.
A sudden change in one or some of these parameters may indicate an abnormal behavior.
In this work, an \flo{entire} day (24 hours) is considered as the \flo{period of the time interval}.

Upon detection of abnormal behavior, the events sequence and parameters are checked against known failure patterns to conclude the final decision about a potential failure occurrence.
To perform this verification, the detection mechanism must be aware of the time, nodeID, and type of occurring events.
\flo{This information is} provided via the \emph{timestamp}, \emph{location}, and \emph{message} fields of the syslog entries.
Additional information, such as usernames and IP addresses, enhance the accuracy of analysis.
According to \sia{the general privacy} guidelines this additional information must remain private.

The message field of the syslog entries also contains detailed information about the users and system behavior, as well as sensitive information such as usernames.
\flo{Therefore, anonymization must be performed before failure detection and identification.}
$PaRS$ is herein used to anonymize the Taurus system logs.

The detection mechanism \flo{involves} two main tasks in view of understanding system behavior via syslog entries.
First, the identification of similar events and second, the distinction of the differences between similar patterns.
For example the log entries \#2 and \#9 from~\autoref{tab:sample-logs} report the occurrence of similar events.
\flo{However, different users triggered each of these similar events.}
For the detection mechanism it is important to understand that the same \flo{type} of event \flo{occurred} by different users of the system.
The \emph{category} in~\autoref{tab:sample-logs}, is essentially the event pattern of each system log entry, and enables the detection mechanism to identify similarities between events.
The hash key is employed for representing the differences.
Therefore, choosing a proper form of de-identification plays a key role in the accuracy of the failure detection mechanism.
A de-identification form close to global de-identification, radically generalizes events with various properties into a single hash key.
Conversely, a de-identification form close to individual de-identification, generates unnecessary degrees of differences between similar events.

\section{Assessing Data Usefulness}
\label{the-usefulness}
Data \emph{usefulness} can only be evaluated \flo{with respect to} a certain intended usage.
In this work, the intended use of the anonymized system logs is failure \flo{detection and identification} on Taurus high performance computer.

Recall the five types of information in each syslog entry:
(1)~\emph{time} of the event,
(2)~\emph{location} of the event,
(3)~\emph{action} which has been done,
(4)~user that triggered the event,
and (5)~detailed information about the action.
Some of this information must be removed to ensure the data privacy, and some must remain to preserve the usefulness of the syslog entry.
The information which has to be removed from log entries depend on the privacy guidelines of the system.
However, for the purpose of failure analysis, regardless of the type of system log entry and the administration domain, (1)~\emph{time}, (2)~\emph{location}, and (3)~\emph{action} of the event must be kept in order to preserve the basic semantic of the syslog entries.

In~\autoref{lst:singleLogEntry} (Section~\ref{sec:encoding}), user could be any of the registered system users.
The detailed information about the action could be any available path/command in the system.
Note that not all paths/commands in a system are executable and also not all of the executed paths/commands \flo{appear} in system log entires.
The same \flo{rationale} applies to the users of a system. 
\flo{For example, not all users have access to all executable commands that appear in system log entires.}
This assumption is valid for all variable terms in system log entries.
Therefore, we can consider two separate domains for each variable terms in system log entries: a \emph{theoretical} domain and a \emph{practical} domain.
The theoretical domain of a variable term describes all possible values for that particular term, while the practical domain expresses the values which have been already assigned to that term in the syslog dataset.
The theoretical domain for usernames in the sample syslogs of~\autoref{tab:sample-logs} includes all defined usernames on the computing system.
In contrast, the practical domain only covers the usernames: \emph{siavash}, \emph{florina}, and \emph{root} which appear in the current system log entries.
The parameter $N^{val}_v$ denotes the number of possible values for a significant variable term $v$ according to its practical domain in a raw system log entry $e_r$.

For the log entry \#3 from~\autoref{tab:sample-logs-2}, the only significant variable term is the username, which in the provided dataset took three different values: \emph{siavash}, \emph{florina}, and \emph{root}.
Therefore, the number of values, $N^{val}_{v_1}$ where $v_1$ is username, of entry \verb|(siavash) CMD (#PATH#)| is 3.
This indicates that there are three unique syslog entries in the syslog dataset shown in~\autoref{tab:sample-logs-2} with an identical event pattern.

As a result of generalization of syslog entries via de-identification, the diversity of de-identified log entries is always equal to or less than the diversity of raw log entries.
As~\autoref{tab:sample-logs} illustrates, a complete de-identification of 20 sample log entries converted 15 unique events into 6 event categories (event patterns).
In this example, 10 events belong to category \verb|66dc2742|, while each of the remaining 5 event categories contains 2 events.

Parameter $D_{p} = {f_{p}}/{N^d_p}$ describes the dominance degree of event pattern $p$ in a set of encoded syslog entries.
For the sample data given in~\autoref{tab:sample-logs}, $D_{66dc2742} = {10}/{20}$.
Therefore, the event pattern \verb|66dc2742| has a degree of dominance of $50\%$ in the encoded syslog entires in~\autoref{tab:sample-logs}, while each of the other event patterns hold a $10\%$ degree of dominance.
Considering this assumption, the data usefulness ($U$) is defined as follows:
\begingroup
\fontsize{8.5pt}{10pt}\selectfont
\begin{equation}
\label{eq:Usefulness}
U = \sum_{p=1}^{P} D_{p}\cdot\left(\frac{\prod_{i=1}^{n} N^{s}_{v_i}}{\prod_{i=1}^{n} N^{val}_{v_i}}\right)
\end{equation}
\endgroup

% !TEX root =  ispdc2018.tex
\section{Evaluation}
\label{sec:evaluation}
To evaluate the data usefulness from anonymized system logs, failure detection is performed both, on de-identified and encoded system logs, and the results are compared.
The mechanism for failure detection detects abnormal node-level behavior by monitoring the individual computing nodes'\footnote{Taurus HPC cluster consists of 2046 nodes.} behavior and comparing it either to the previous known \emph{normal} behavior or to the behavior of the majority.
Each node generates entries in the syslog for certain events.
Examples include user login/logout, file-system alerts, authentications.
The frequency, order of occurrence, and time between events provide an individual \emph{syslog entry generation pattern} to each node over time.
Even though this pattern changes according to the workload on the node and other environmental parameters, the pattern mostly remains similar to previous patterns of the same node.
Previous studies showed that the majority of neighboring nodes (located in a similar rack or island) also exhibit similar syslog generation patterns~\cite{Ghaisvand2016a}.
To extract the patterns, a failure detection mechanism needs to assign an event category to each syslog entry.
For example, the same event category will be assigned to all three syslog entries shown in~\autoref{lst:eventtypes}, since all indicate the occurrence of a similar event (execution of a command by a user), even though each user executed a different command.

\begin{lstlisting}[title=SS,language=C,caption=Sample syslog entries of the same event category,label=lst:eventtypes,frame=none,basicstyle=\scriptsize\ttfamily,keepspaces=true]
1517266801 T-1320 (siavash) CMD (/home/config.sh )
1517266913 T-2013 (florina) CMD (/usr/bin/check)
1517267452 T-3114 (root) CMD (/usr/lib32/cl/cl2 1 1)
\end{lstlisting}

Usernames also need to be de-identified before further analysis.
Therefore, the sample syslog entries from~\autoref{lst:eventtypes} should first be de-identified as shown in~\autoref{lst:anonymized}.

\begin{lstlisting}[title=SS,language=C,caption=De-identified sample syslog entries,label=lst:anonymized,frame=none,basicstyle=\scriptsize\ttfamily,keepspaces=true]
1517266801 T-1320 (#USER1#) CMD (/home/config.sh )
1517266913 T-2013 (#USER2#) CMD (/usr/bin/check)
1517267452 T-3114 (#USER3#) CMD (/usr/lib32/cl/cl2 1 1)
\end{lstlisting}

\begin{figure}
	\begin{center}
		\includegraphics[width=.45\textwidth]{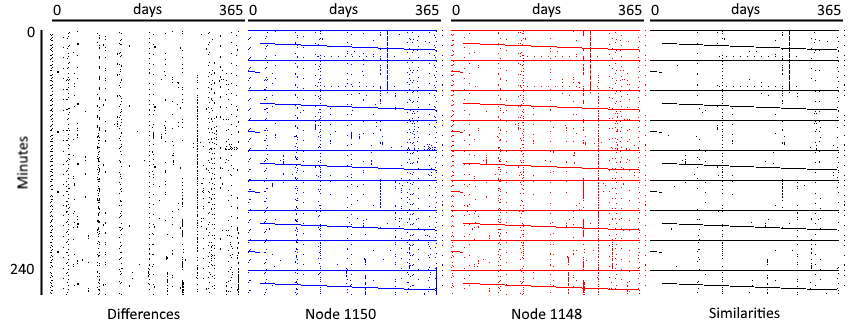}
		\caption{{\em Raw} syslog events generated on nodes 1148 (middle, blue color) and 1150 (middle, red color), between \mbox{2017-01-01} to \mbox{2017-12-31} (leap year) represented as colored dots. 
		Days are indexed from 0 to 365.  
		Time interval shown between 00:00 and 04:00 is indexed from 0 to 240 minutes. 
		White space denotes the absence of an entry on (day, minute). 
		Similarities (left) and differences (right) between the generated syslog events are represented as black color dots.}
		\label{fig:pattern_diff_simi}
	\end{center}
	\vspace{-0.2cm}
\end{figure}

\autoref{fig:pattern_diff_simi} illustrates the partial results of applying the failure detection mechanism on anonymized syslog entries corresponding originating from two Taurus nodes over a period of one year, from January 1, 2017 to December 31, 2017.
The horizontal lines therein illustrate periodic events; the absence of a periodic event is a sign of a potentially abnormal behavior.
The vertical lines in two middle plots, also indicate a potential abnormal behavior,
because of their rare and random occurrence.
The existence of vertical lines at the same place on the right most plot (similarities), decreases the probability that an abnormal behavior occurred.
In contrast, any vertical or horizontal line in the left most plot (differences) is an indicator of a highly probable abnormal behavior.

In the present failure detection pass, a de-identified version of the syslog entries was employed.
With the individual de-identification method only the required sensitive information has been removed and no generalization was applied.
Therefore, the current anonymized dataset, regardless of the number of significant variable terms, has the highest possible data usefulness, namely $100\%$.
Applying generalization on the syslog data via group de-identification or global de-identification decreases, however, the data usefulness.
Equation~\eqref{eq:Usefulness} can be used to calculate the data usefulness.
Considering~\autoref{tab:sample-logs-1} as the sample dataset and \emph{usernames}, \emph{daemons}, and \emph{jobs} number as \emph{significant} variable terms, the data usefulness of the anonymized version of the dataset is calculated as follows.
\autoref{tab:sample-logs-3} illustrates the intermediate values required in Equation~\eqref{eq:Usefulness}.
When a log entry contains no significant variable terms, $n=0$, $\left(\frac{\prod_{i=1}^{n} N^{s}_{v_i}}{\prod_{i=1}^{n} N^{val}_{v_i}}\right)$ is considered to be $1$.

The usefulness of sample data logs in~\autoref{tab:sample-logs-1}, anonymized via group and global de-identification is $78\%$ and $61\%$ respectively.
\begingroup
\fontsize{8.5pt}{10pt}\selectfont
\begin{align*}
U &= \left(0.5\cdot\frac{1}{3}\right)+\left(0.1\cdot\frac{1}{1}\right)+\left(0.1\cdot\frac{1}{1}\right)+\left(0.1\cdot1\right) \\
&+ \left(0.1\cdot\frac{1}{2}\right)+\left(0.1\cdot\frac{1}{1}\right) \\
&\approx 0.616
\end{align*}
\endgroup
The usefulness of data in~\autoref{tab:sample-logs-1} via global de-identification is approximately $61\%$ compared to the raw data.
By dividing the usernames into two groups of normal (\emph{florina} and \emph{siavash}) and privileged (\emph{root}) users, and substituting the username of each user with a unique group identifier (e.g. \verb|#USRn#|, \verb|#USRp#|),  data usefulness increases to $78\%$.
Replacing each username with a unique identifier (e.g., \verb|#USR1#|, \verb|#USR2#|, \verb|#USR3#|), brings the data usefulness back up to $100\%$ for the purpose of failure detection and identification.
It is important to note that the encoding phase has no influence on the data usefulness for the intended purpose.
As shown in~\autoref{tab:sample-logs-2}, each encoded syslog entry is labeled with its event \emph{category} (event pattern hash key).
Therefore, the similarity between entries is detectable even though the entires are encoded.

For the calculation of the data usefulness for the sample syslog entries in~\autoref{tab:sample-logs-1}, the intermediate values shown in~\autoref{tab:sample-logs-2} are needed.
To calculate the intermediate values, the significant variable terms of each event must be identified as well as their practical domains.
In this example, there are only 20 syslog entries and calculating the intermediate values is not challenging.
However, calculating such values for a much larger dataset is challenging.
Therefore, a closer look at the Taurus syslog dataset is needed before calculating the data usefulness for encoded Taurus syslog entries.

The collection of Taurus syslog entries from January to December 2017 consists of more than $3.7$ billion log entries.
$75\%$ of all raw log entries appeared more than once.
After global de-identification of all syslog entries, the $3.7$ billion raw entries resulted into less than $484$ thousand unique event patterns, a very high number of events for calculating the intermediate usefulness values~\eqref{eq:Usefulness}.
Even though, on average, each event pattern should appear about $7,700$ times in the entire syslog entry dataset, further analysis revealed that the share of event patterns is very different.
More than $71\%$ of all syslog entries in Island 1 of Taurus HPC cluster were derived from only $5$ unique event patterns.
And only the $50$ most frequent event patterns \flo{were derived from} $3.46$ billion ($93.65\%$) syslog entries.

\autoref{tab:top-patterns} denotes the percentage of events generated by the top 5, 25, and 50 \emph{most frequent patterns} in comparison with the total number of events on each island and also across all islands.
The first and second rows of the table show the total number of raw syslog entries and the number of event patterns as a reference.
In each island, $92\%$ to $97\%$ of the events match fewer than 50 unique event patterns, i.e., $0.01\%$ of all event patterns.

\begin{table}[!t]
	\fontsize{5.5pt}{9pt}\selectfont
	\caption{Percentage of syslog entries that yield the most frequent patterns (M--millions, K--thousands)}
	\centering
	\begin{tabular}{@{}r@{}| r | r| r | r | r| r | r}
	\textbf{ }& \textbf{Island 1}& \textbf{Island 2}& \textbf{Island 3}& \textbf{Island 4}& \textbf{Island 5}& \textbf{Island 6}& \textbf{Total}\\
	\hline
	\#Raw log entries&420 M &59 M&203 M&403 M&1291 M&1336 M&3714 M\\
	\#Event patterns&225 K&12.6 K&27.7 K&65.5 K&65.6 K&157 K&484 K\\
	Top 5&71.49\%&43.96\%&64.13\%&25.65\%&33.06\%&29.38\%&28.76\%\\
	Top 25&94.46\%&88.38\%&95.02\%&78.71\%&80.58\%&81.06\%&76.47\%\\
	Top 50&97.00\%&92.15\%&97.17\%&94.78\%&94.31\%&95.08\%&93.65\%\\
	\end{tabular}
	\label{tab:top-patterns}
\end{table}

Based on this observation, the focus can be on the most frequent event patterns that match the majority of the system log entries.
The significant variable terms should be also identified according to the intended data usage.
The $5$ most frequent event patterns for Island~1 of Taurus, are shown in~\autoref{lst:top5}.
These $5$ event patterns match $71\%$ of all system logs generated on Taurus-Island~1.

\begin{lstlisting}[title=SS,language=C,caption=Five most frequent event patterns and their frequency,label=lst:top5,frame=none,basicstyle=\scriptsize\ttfamily,keepspaces=true]
43% (#USR_#) CMD (#PATH#)
 9% starting #DAEM#
 9% finished #DAEM#
 5% Received disconnect from #IPv4# disconnected by user
 5% pam_unix(sshd:session): session closed for user #USR_#
\end{lstlisting}

Usernames (\#USER\_\#) and IP addresses (\#IPv4\#) must be anonymized for data privacy.
Different system paths (\#PATH\#) do not affect the nature of an event with respect to the failure detection and identification, therefore, they are also considered as non-significant variable terms.
Daemon names (\#DAEM\#) carry important information and greatly affect the meaning of events recorded within syslog.
There are 14 daemon names in the syslog dataset under consideration.
Based on the definition of significant variable terms for the $5$ most frequent event patterns, the data usefulness can be calculated as follows.
Parameter $A$ represents the quality of the syslog entries that do not originate in the 5 most frequent patters and which correspond to $28\%$ of the total syslog entries.
\begingroup
\fontsize{8.5pt}{10pt}\selectfont
\begin{align*}
U &= (0.43\cdot1)+\left(0.09\cdot\frac{1}{14}\right)+\left(0.09\cdot\frac{1}{14}\right)+(0.05\cdot1) \\
&+ (0.05\cdot1)+A \\
&\approx 0.542 + A
\end{align*}
\endgroup
\begin{figure}[!h]
	\begin{center}
		\includegraphics[width=.45\textwidth]{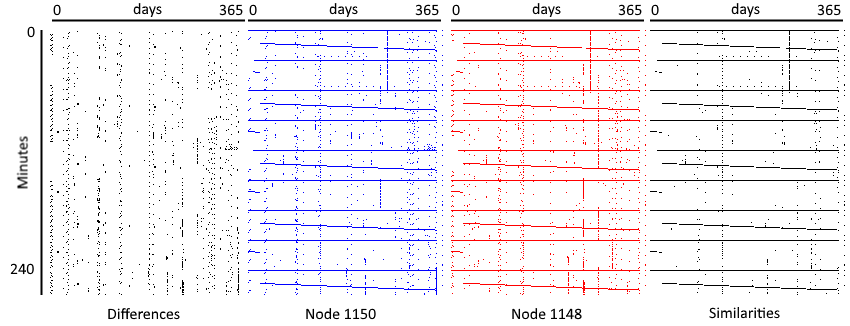}
		\caption{{\em Encoded} syslog events generated on nodes 1148 (middle, blue color) and 1150 (middle, red color), between \mbox{2017-01-01} to \mbox{2017-12-31} (leap year) are represented as colored dots. Days are indexed from 0 to 365.  Time interval shown between 00:00 and 04:00 is indexed from 0 to 240 minutes. White space denotes the absence of an entry on (day, minute). 
		 Similarities (left) and differences (right) between the generated syslog events are represented as black color dots.}	
		\label{fig:pattern_diff_simi_2}
	\end{center}
	\vspace{-0.2cm}
\end{figure}

Considering the $50$ most frequent event patterns, improves the accuracy of data usefulness calculation.
None of the next $45$ most frequent event patterns have significant variable terms, therefore $x=1$ and the final value of usefulness, even by assuming $0$ utility for the remaining $1\%$ of event patterns, is as high as $82.2\%$.
Despite of the fact that there is a certain degree of information loss, the visual illustration of the results shown in~\autoref{fig:pattern_diff_simi_2} also confirms the high quality of encoded system logs for the intended usage of failure detection and identification.
\begingroup
\fontsize{8.5pt}{10pt}\selectfont
\begin{align*}
U &= (0.43\cdot1)+\left(0.09\cdot\frac{1}{14}\right)+\left(0.09\cdot\frac{1}{14}\right)+\left(0.05\cdot1\right) \\
&+ (0.05\cdot1)+(0.28\cdot x)+(0.01\cdot0) \\
&\approx 0.822
\end{align*}
\endgroup
% !TEX root =  ispdc2018.tex
\section{Conclusion and Future work}
\label{sec:conclusion}
This work introduced an approach for assessing the impact of encoding, for the purpose of anonymization, on the \emph{usefulness} of data in the analysis of failures on HPC systems.
Due to the presence of a large number of personal identifiers in system logs, anonymization of syslog data is a prerequisite for their \mbox{regulation-compliant} 
processing and sharing~\cite{GDPR}.
The proposed anonymization approach strikes a balance between data privacy and data usefulness in a manner that corresponds to the goal of the data analysis (e.g., analysis of failures reported in HPC system logs).
The distinctive aspect of the proposed anonymization approach is employing irreversible encoding via a collision-resistant hashing function in the final step of the anonymization process.
This guarantees data privacy.
A~metric to quantify the usefulness of the anonymized data has been introduced that is tailored to the intended data analysis method for failure identification. 
This ensures data usefulness.
Moreover, due to the encapsulation of long entries from the raw system log into short constant-length entries in the encoded system log, up to $90\%$ saving on storage space is possible.

This work also showed that on a production HPC system, less than $1\%$ of log event patterns match more than $92\%$ of all syslog entries.
Therefore, the proposed approach for assessing the data usefulness is feasible and of high practical value.
It has been also shown that the encoded system logs, even after full anonymization via the proposed anonymization approach, can be used for the purpose of failure detection and analysis.
Adjusting the usefulness parameter, encoding and de-identification of time and location terms from the log entries, as well as adapting the proposed approach for other types of system logs are planned as part of future work.

\section*{Acknowledgement}

This work is in part supported by the German Research Foundation (DFG) within the Cluster of Excellence `Center for Advancing Electronics Dresden (cfaed)', and by Eucor~--~ \mbox{The European} Campus, within the Seed Money project `Data Analysis for Improving High Performance Computing Operations and Research'.

\begin{spacing}{0.92}
{\small
\Urlmuskip=0mu plus 1mu\relax
\bibliographystyle{./style/IEEEtranBST/IEEEtran}
\bibliography{literature}
}
\end{spacing}

\end{document}